\documentclass[twocolumn]{aastex631}

\usepackage{siunitx}
\usepackage{bm}
\usepackage{mathtools}
\usepackage{multirow}
\usepackage{CJK}         
\usepackage{hyperref}
\defcitealias{Secunda:2024}{Paper~I}

\begin{document}
\begin{CJK*}{UTF8}{gbsn}

\title{Continuum Reverberation in Active Galactic Nuclei Disks Only With Sufficient X-ray Luminosity and Low Albedo}

\author[0000-0002-1174-2873]{Amy Secunda}\thanks{E-mail: asecunda@flatironinstitute.org}
\affil{Center for Computational Astrophysics, Flatiron Institute, New York, NY 10010, USA}

\author[0000-0002-2624-3399]{Yan-Fei Jiang (姜燕飞)}
\affil{Center for Computational Astrophysics, Flatiron Institute, New York, NY 10010, USA}

\author[0000-0002-5612-3427]{Jenny E. Greene}
\affil{Department of Astrophysical Sciences, Princeton University, Peyton Hall, Princeton, NJ 08544, USA}

\begin{abstract}
Disk continuum reverberation mapping is one of the primary ways we learn about active galactic nuclei (AGN) accretion disks. Reverberation mapping assumes that time-varying X-rays incident on the accretion disk drive variability in UV-optical light curves emitted by AGN disks, and uses lags between X-ray and UV-optical variability on the light-crossing timescale to measure the radial temperature profile and extent of AGN disks. However, recent reverberation mapping campaigns have revealed oddities in some sources such as weakly correlated X-ray and UV light curves, longer than anticipated lags, and evidence of intrinsic variability from disk fluctuations. To understand how X-ray reverberation works with realistic accretion disk structures, we perform 3D multi-frequency radiation magnetohydrodynamic simulations of X-ray reprocessing by the UV-emitting region of an AGN disk using sophisticated opacity models that include line opacities for both the X-ray and UV radiation. We find there are two important factors that determine whether X-ray irradiation and UV emission will be well-correlated, the ratio of X-ray to UV luminosity and significant absorption. When these factors are met, the reprocessing of X-rays into UV is nearly instantaneous, as is often assumed, although linear reprocessing models are insufficient to fully capture X-ray reprocessing in our simulations. Nevertheless, we can still easily recover mock lags in our light curves using software that assumes linear reprocessing. Finally, the X-rays in our simulation heat the disk, increasing temperatures by a factor of 2--5 in the optically thin region, which could help explain the discrepancy between measured and anticipated lags.
\end{abstract}

\section{Introduction}
\label{ch5:sec:into}

Active galactic nuclei (AGN) accretion disks are important laboratories to study accretion physics and fuel powerful radiation and outflows that are thought to have a large impact on galaxy formation \citep{Fabian2012,Kormendy2013}. Unfortunately, these AGN accretion disks are generally too distant to resolve spatially. However, because the emission from the AGN corona, accretion disk, and broad line region vary with time, temporal resolution can be substituted for spatial resolution using a technique known as reverberation mapping.

Reverberation mapping was first proposed by \cite{Blandford1982} as a way to measure distances to the broad line region by measuring the time lag between variability in continuum light curves and broad lines \cite[see also,][]{Kaspi_2000,Peterson2004,Bentz_2015,Grier:2017}. \cite{Collier:1999} later introduced the idea of disk continuum reverberation mapping. Disk continuum reverberation mapping is often interpreted in the context of the lamp-post model, which proposes that time-varying X-ray radiation emitted by the corona will move outwards to different temperature regions of the accretion disk where it will be absorbed, reprocessed, and re-emitted at different wavelengths based on the local temperature of the disk. Because X-ray radiation takes a light-crossing timescale to travel from the inner hotter regions of the disk to the outer cooler regions, we can use the lag between variability in longer wavelength bands and variability in shorter wavelength bands to measure the radial temperature profile and radial extent of an AGN disk \citep[e.g.,][]{Sergeev:2005,Cackett:2007,Cackett:2018,derosa2015,Edelson:2015,Edelson2017,Edelson:2019,Jiang:2017,Fausnaugh:2016,Starkey:2017,Homayouni:2022}.

Traditionally, the lamp-post model assumes that the X-ray variability is the main driver of variability in UV-optical light curves. However, there are a growing number of examples of X-ray and UV-optical light curves that are only moderately \citep[e.g.,][]{Edelson:2019,Cackett:2023,Kara:2023} or weakly \citep{Schimoia:2015,Buisson:2018} correlated. This lack of correlation could be due to absorption by intervening material \citep[e.g.,][]{Kara:2021} or changes in the height and temperature of the corona on timescales shorter than the observing timescale \citep{Panagiotou:2022B}. Alternatively, there is observational evidence for intrinsic variability in UV-optical light curves emitted by AGN disks \citep[e.g.,][]{Arevalo:2009,Neustadt:2022,F92020,Beard:2025}. This variability could arise from fluctuations in the UV-optical regions of the accretion disk due to magnetorotational instability (MRI) driven turbulence \citep{BalbusHawley1991} or convection resulting from the enhanced opacity in the UV-optical region of the disk \citep{JiangBlaes2020}. 

The recent development of accurate, high-speed, numerical methods to calculate radiative transfer coupled to magnetohydrodynamics (MHD) makes it possible for the first time to examine how both radiation and magnetic fields have an impact on AGN disk structure and turbulence \citep{Jiang2013,JiangBlaes2020,Jiang2021,Hopkins:2023}.  However, these MHD simulations model radiation using a single frequency group, with the radiation integrated over all frequencies, and do not examine the impact of X-ray irradiation on the disk and the light curves emitted by the disk. Instead, reverberation mapping mostly employs empirical models to describe the reprocessing of X-ray irradiation into UV-optical emission. 

The most common of these empirical models is a linear reprocessing model. In a linear reprocessing model the reprocessed light curve is,
\begin{equation}
\label{eq:ch5_reproccess}
        \rm r(t) = \int_{-\infty}^{\infty}\psi(\tau_{\rm L})d(t-\tau_{\rm L})d\tau_{\rm L},
\end{equation}    
where $\tau_{\rm L}$ is the continuum lag on the light-crossing timescale, $d(t)$ is the driving light curve, and $\psi(t)$ is the response function, which is commonly modeled as a Gaussian or log-normal function. There have been efforts to develop more sophisticated theoretical models of reverberation mapping that can help explain the discrepancies between the lamp-post model and observations. However, these models are primarily (semi-)analytic and make assumptions about the vertical profiles of various disk parameters, which often remain fixed over time or are not calculated self-consistently to include the interaction of radiation and gas \citep[e.g.,][]{Sun:2020,Kammoun:2021b,Salvesen:2022,Panagiotou:2022,Best:2024,Ren:2024}.
 
In \citet{Secunda:2024}, hereafter \citetalias{Secunda:2024}, we provided an early application of a multi-frequency radiation MHD code described in \cite{Jiang2022}, which instead of integrating over all frequencies of light, allows the user to integrate over different frequency bands. In \citetalias{Secunda:2024}, we used two frequency groups to simulate the reprocessing of high-frequency or X-ray radiation into low-frequency or UV radiation by the AGN disk. Unlike in the previous semi-analytic models, in these simulations the radiation energy and momentum are coupled to the gas evolution, and the temperature, opacity, and energy of the simulation depend self-consistently on the local gas density and pressure. We found that fluctuations in the disk due to the MRI and convection lead to intrinsic variability in the UV light curves emitted by the AGN disk, and the power spectral density (PSD) of this variability is similar to observed UV-optical AGN light curve PSDs. We also found that due to the low absorption opacity for the X-ray radiation, the injected X-ray and reprocessed UV light curves from the simulations were only weakly correlated. 

\citetalias{Secunda:2024} used tabulated Rosseland and Planck mean opacities that include line opacities in addition to electron scattering and free-free absorption for the low-frequency radiation group, because \citep{JiangBlaes2020} showed that including line opacities for UV radiation has a large impact on AGN disk models. However, the opacities for the high-frequency group were modeled simply using the proper average of the free-free absorption opacity and electron scattering opacity. 

In this paper, we build on \citetalias{Secunda:2024} by using a more sophisticated opacity model for the X-ray component in order to see if with more realistic opacities the X-ray absorption will be enhanced, increasing the correlation between the injected X-ray and reprocessed UV light curves. A higher correlation between the injected X-ray and reprocessed UV light curves, would allow us to test the linear reprocessing model commonly used to describe reverberation mapping.

We summarize our simulation set-up, highlighting the differences between the simulations in \citetalias{Secunda:2024} and this paper in Section \ref{ch5:sec:methods}. We describe the disk structure of our simulations and present the light curves from our different runs in Sections \ref{sec:ch5_disk} and \ref{sec:ch5_lcs}, respectively. Because we find that our new opacity model for the X-ray radiation leads to strongly correlated X-ray and UV light curves in several of our simulations, we are able to test different models of the response function for the reprocessing of X-ray irradiation into UV emission on our simulations for the first time in Section \ref{sec:ch5_response} and also examine the performance of {\sc javelin}, a commonly used lag detection method first developed by \cite{Zu2011}, on our simulated light curves in Section \ref{sec:ch5_javelin}. Finally, we summarize and discuss these updated results in Section \ref{ch5:sec:discuss}.

\section{Methods}
\label{ch5:sec:methods}

\begin{table}[]
    \begin{tabular}{c|c}
        Parameter & Unit \\
        \hline \\
        \bm{$\Omega_0$} & $\num{4.73e-6}$~s$^{-1}$\\
        \bm{$\rho_0$} & $10^{-7}$~g~cm$^{-3}$ \\
        \bm{$T_0$} & \num{2e5}~K \\
        \bm{$H_{\rm g}$} & $\num{1.11e12}$~cm \\
        $H_{\rm r}$ & 4.06~$H_{\rm g}=\num{4.51e12}$~cm \\
        $r_0$ & $r_0=45$~R$_{\rm s}=\num{6.68e14}$~cm \\
        $\tau_0$ & $\num{6.26e4}$ \\
        $P_{\rm g,0}$ & \num{2.77e6}~dyn~cm$^{-2}$ \\
        $\beta_{\rm x,y}$ & 200 \\
        $\beta_{\rm z}$ & $2 \times 10^4$ \\
        $P_{\rm rat}=P_{\rm r,0}/P_{\rm g,0}$  & 4.37 \\
        $c_{\rm rat}=c/c_{\rm s}$  & \num{5.69e3}\\
        $N_x$ & 48 \\
        $N_y$ & 192 \\
        $N_z$ & 1536 \\
        $L_x$ & 3~$H_{\rm g}$ \\
        $L_y$ & 12~$H_{\rm g}$ \\
        $L_z$ & 96~$H_{\rm g}$ \\
        $N_{\theta}$ & 48 \\
        $\tau_{\rm therm}$ & 32~days\\
    \end{tabular}
    \caption{$\Omega_0$, $\rho_0$, $T_0$, and $H_{\rm g}$ are fiducial units we use as normalization parameters in our simulation for time, mass, temperature, and distance, respectively. $H_{\rm r}$ is the initial total (gas and radiation combined) pressure scale height. $r_0$ is the fixed location of the center of the shearing box and $\tau_0$ and $P_{\rm g,0}$ are the initial midplane optical depth and gas pressure, respectively. $\beta_{\rm x,y}$ and $\beta_{\rm z}$ are the ratio of the gas to magnetic pressure for the zero net-flux and vertical magnetic fields, respectively. $P_{\rm rat}$ and $c_{\rm rat}$ are the initial ratio of the radiation pressure to the gas pressure and the ratio of the speed of light to the sound speed, respectively, at the midplane. $N_x$, $N_y$, $N_z$ are the number of cells in the $x$, $y$, and $z$ direction, while $L_x$, $L_y$, $L_z$ are the dimensions of the box. $N_{\theta}$ is the number of discrete angles in each cell for each frequency group. Finally, $\tau_{\rm therm}$ is the approximate thermal timescale calculated from the average $\alpha$-viscosity in our simulations once they reach steady-state.}
    \label{tab:ch5_units}
\end{table}

\begin{table*}[]
\centering
    \begin{tabular}{c|c}
       Run  &  X-ray Light Curve Parameters\\
       \hline
       Fiducial  & DRW PSD with $\tau_{\rm damp} = 30$~days \\
       Low Flux  & DRW PSD with $\tau_{\rm damp} = 30$~days, 1/4 flux of Fiducial \\
       Power-law & Power-law PSD with $P(\nu) \propto \nu^{-2}$ \\
       Shallow Power-law & Power-law PSD with $P(\nu) \propto \nu^{-1.5}$
    \end{tabular}
    \caption{A summary of the different properties of the X-ray light curve injected into the simulation for the four different runs in this paper.}
    \label{tab:ch5_runs}
\end{table*}

\begin{figure}
    \includegraphics[width=\columnwidth]{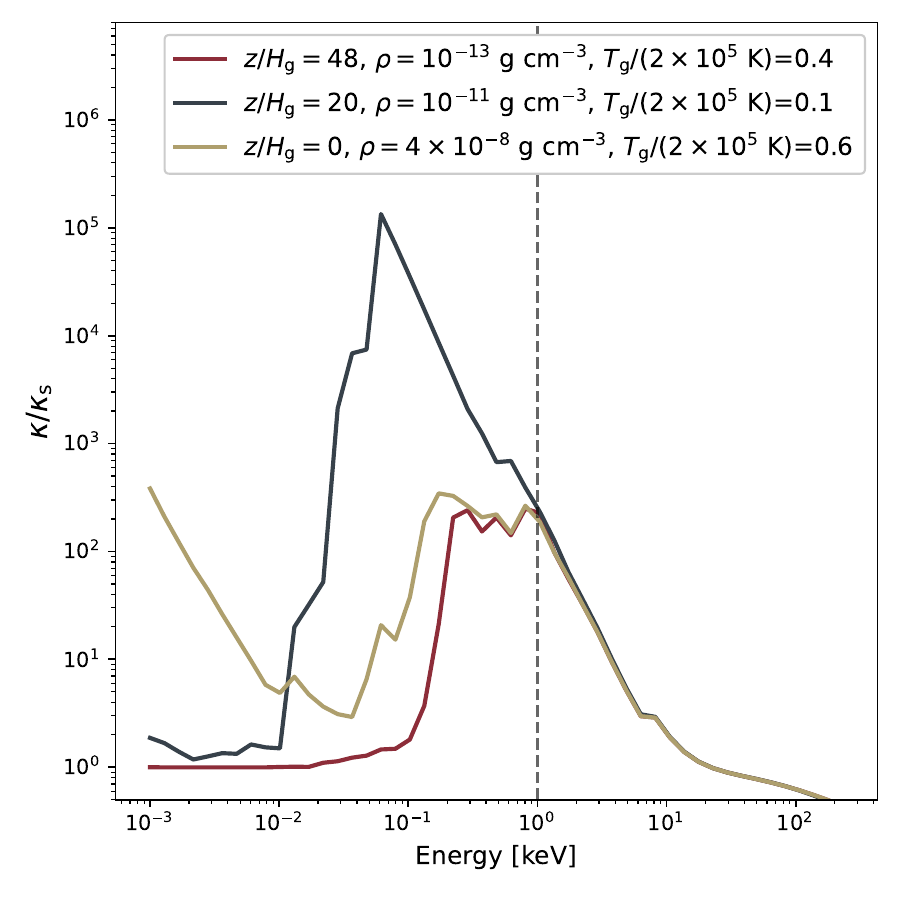} \\
    \includegraphics[width=\columnwidth]{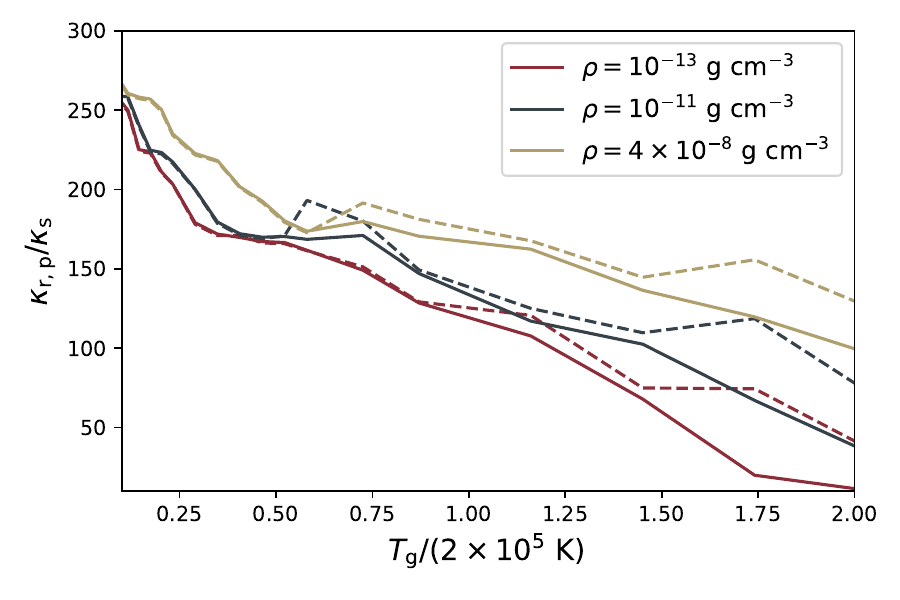}
    \caption{\textbf{Top panel:} The opacity, $\kappa$, scaled by the scattering opacity, $\kappa_{\rm s}$, as a function of frequency at densities and temperatures representative of three different heights in the simulation given in the legend. The dashed line shows the cut off between our high- and low-frequency radiation groups at 1~keV. \textbf{Bottom panel:} The Rosseland (solid line) and Planck (dashed line) mean opacities as a function of gas temperature for the X-ray radiation at the three densities in the top panel. While the opacity is greater at lower frequencies, the mean opacities are weighted such that the opacity at the lowest energy in each group dominates. As a result, the mean opacity for the high-frequency group is larger throughout the disk.}
    \label{fig:ch5_kappa}
\end{figure}

The initial set-up of our simulation is similar to \citetalias{Secunda:2024}. In this section, we provide a brief summary of the simulation set-up, highlighting the main differences from \citetalias{Secunda:2024}, and refer the reader to \citetalias{Secunda:2024} for full details.

Using {\sc athena++}, we simulate a local patch of an AGN accretion disk initialized in radiation hydrostatic equilibrium using the local shearing box approximation. In the shearing box approximation the $x$- and $y$-coordinates correspond to the radial and azimuthal dimensions, while the $z$-coordinate corresponds to the height. We use the same parameters as in \citetalias{Secunda:2024} which are summarized in Table \ref{tab:ch5_units}. We use periodic and shear periodic {\sc athena++} boundary conditions for the $x$- and $y$-boundaries, respectively, and outflow boundary conditions for the gas at the $z$-boundary. We initialize the magnetic field in our simulation as in \citetalias{Secunda:2024} using two oppositely twisted flux tubes with zero net-flux and an additional purely vertical magnetic field to increase the Maxwell stress.

We solve the implicit radiative transfer equations for intensities over discrete angles coupled to the MHD solver in {\sc athena++} for two frequency groups in the manner described in \cite{Jiang2021} and \cite{Jiang2022}. We use the same two frequency groups as in \citetalias{Secunda:2024}, one group covering the frequency range $[0,1\text{keV}]$ to model the radiation field emitted by the disk locally (at UV wavelengths) and the other group covering the frequency range $(1\text{keV},\infty)$ for the X-ray irradiation. We initialize the UV radiation isotropically based on the local disk temperature. We initialize the X-ray radiation isotropically to a constant radiation density four orders of magnitude weaker than the midplane radiation energy for the UV frequency group, as in \citetalias{Secunda:2024}. At the $z$-boundaries, we allow intensities for both frequency groups to leave the simulation domain, with outgoing intensities copied from the last active zone to the ghost zone. For the low frequency group, intensities are not allowed to enter. For the high frequency group, we set the incoming intensities based on a prescribed X-ray light curve. 

Our implicit radiative transfer scheme allows us to calculate the opacity for each frequency group based on the time-evolving gas density and temperature. As in \citetalias{Secunda:2024}, we determine this opacity for the UV component in each cell with a bi-linear interpolation of the {\sc opal} opacity tables from \cite{IglesiasRogers1996} for each time-step based on each cell's current gas temperature and density. The largest change between our previous simulations in \citetalias{Secunda:2024} and this paper is the way we model the opacity for our X-ray component. In \citetalias{Secunda:2024}, we modeled the X-ray opacity as the proper average of the free-free absorption opacity and electron scattering opacity. We found that for the temperatures and densities of our simulation the proper average was dominated by the scattering opacity, which prevented the X-rays in our simulation from being absorbed and led to a low correlation between our injected X-ray and reprocessed UV light curves.

However, much like using tabulated {\sc opal} opacities is crucial for modeling the structure of and fluctuations driving variability in AGN disks \citep[e.g.,][]{Jiang2016}, including a more sophisticated model for opacity for our higher frequency group could be the key to increasing X-ray absorption and the correlation between the driving X-ray and reprocessed UV light curves. We therefore use the online software {\sc tops} \citep{tops}\footnote{\url{https://aphysics2.lanl.gov/apps/}} to calculate for solar composition the Rosseland and Planck mean opacities as a function of density and temperature for photon energies above 1~keV. We show the opacity as a function of frequency for densities and gas temperatures representative of three different heights in our simulation disk in Figure \ref{fig:ch5_kappa}. We normalize the opacity by the scattering opacity $\kappa_{\rm s} =0.3$~cm$^2$~g$^{-1}$, which was the dominant source of opacity in \citetalias{Secunda:2024} for the high-frequency component. 

It is clear from Figure \ref{fig:ch5_kappa} that even at photon energies greater than 1~keV, using mean opacities calculated from {\sc tops} increases the X-ray absorption opacity significantly. In fact, for the temperature range in our simulations the calculated mean Rosseland and Planck opacities are weighted such that the opacity for the lowest photon energies dominate because the Blackbody spectrum peaks at around 0.01~keV for temperatures around $10^5$~K. As a result, the X-ray opacity is actually greater than the UV opacity throughout the simulated disk. We therefore expect this change in opacity model to have a large impact on our simulated light curves.

We perform four different runs which are summarized in Table \ref{tab:ch5_runs}. The only prescribed differences between these runs lies in the way we model the X-ray light curve we inject into the simulation. The first of these runs is our Fiducial run, which uses the same injected X-ray light curve as Run A in \citetalias{Secunda:2024}, a damped random walk (DRW) light curve with $\tau_{\rm damp}=30$~days and mean flux equal to the mean UV flux emitted in our simulation without injected X-rays from \citetalias{Secunda:2024}. We do this in order to directly compare the impact of our new opacity model for the X-ray radiation. In addition, DRW light curves are frequently used to model AGN light curves. Next, we perform a run with the same injected X-ray light curve as the Fiducial run, but with $1/4$ the flux, in order to study the impact of the amount of X-ray flux on reprocessed UV light curves. 

Our next two runs use a power-law instead of a DRW to model the PSD of the injected X-ray light curve. The first, our Power-Law run, has a power-law index of $-2$, which is the same as the power-law component of the DRW, while the second, our Shallow Power-Law run, has a slightly shallower power-law index of $-1.5$. These power-law indices were chosen based on empirical power-law fits to X-ray AGN light curves \citep[e.g.,][]{Markowitz:2003}. 

All X-ray light curves are generated with a timestep of $10^{-3}$~days and a duration of $10^4$ days using the methods in \citet{Kelly2009} and \citet{Timmer1995} for the DRW and power-law light curves, respectively, which assume Gaussian-distributed light curves. Finally, we use for comparison the results from our one-frequency run, or Gray run, in \citetalias{Secunda:2024}, where we do not inject an X-ray light curve and integrate the radiation over all frequencies of light instead of using multiple frequencies groups.

\section{Results}
\label{ch5:sec:results}

\subsection{Disk Structure}
\label{sec:ch5_disk}

\begin{figure*}
    \centering
    \includegraphics[width=0.38\textwidth]{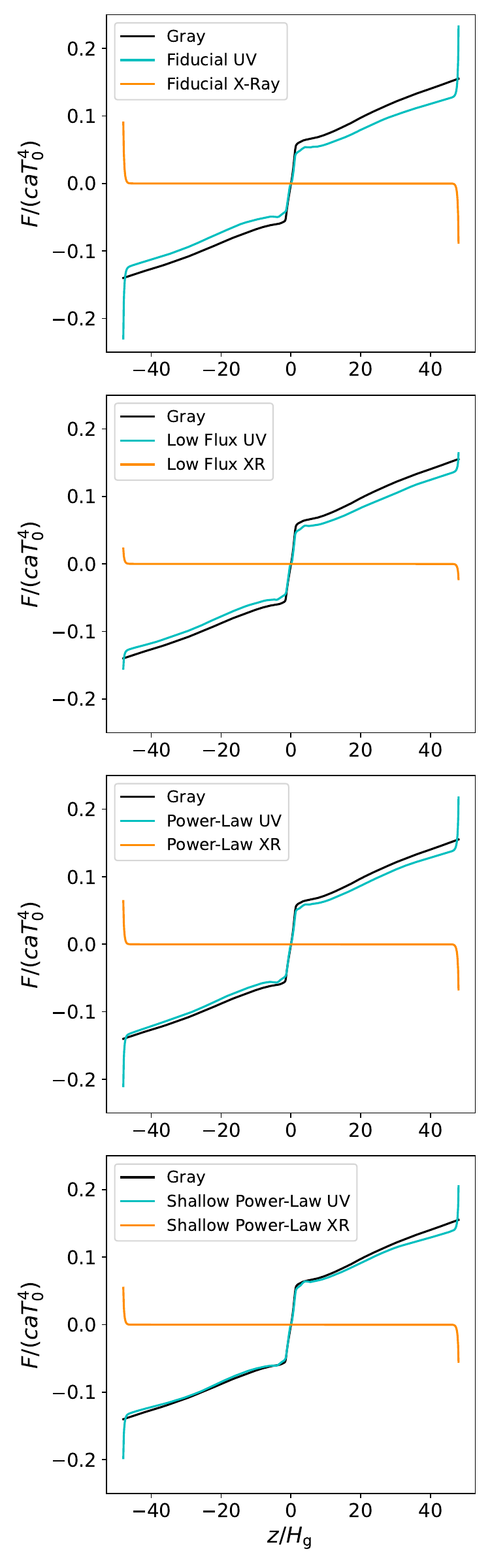} 
    \includegraphics[width=0.45\textwidth]{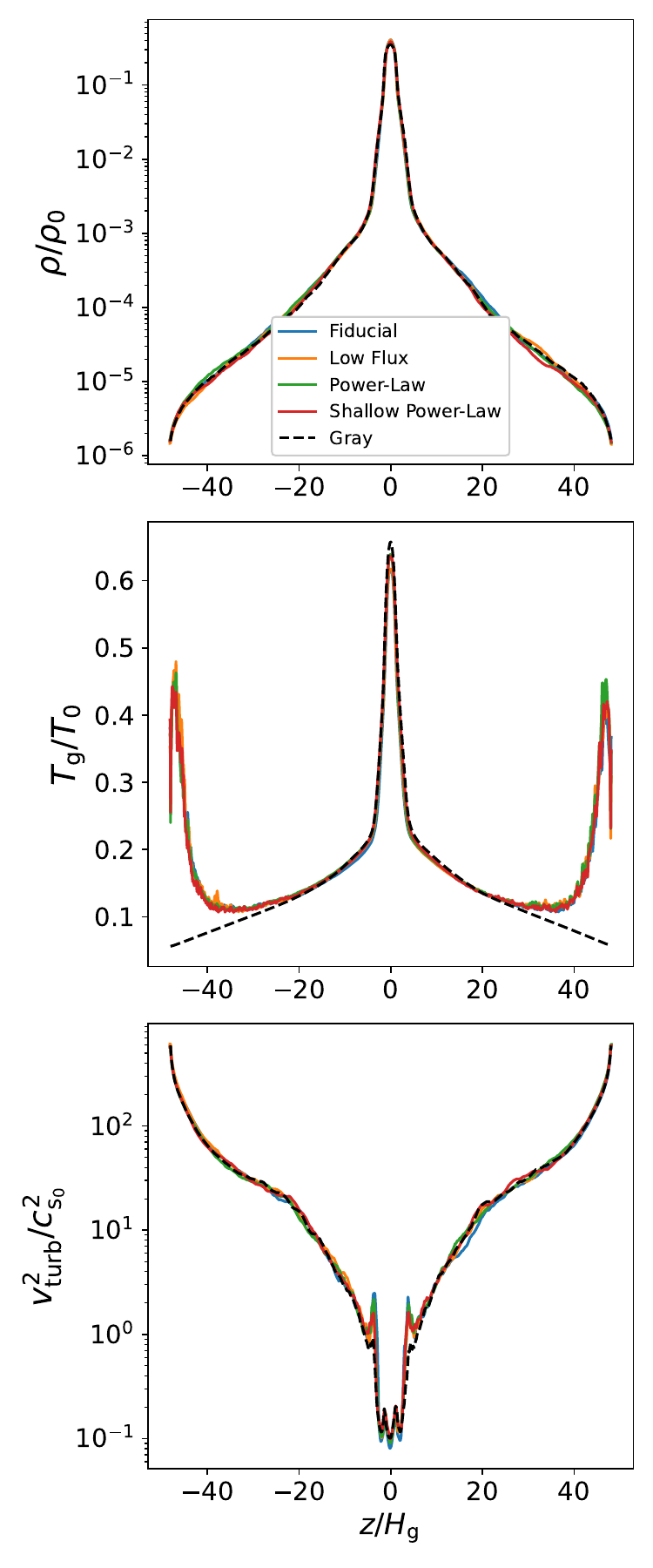}
    \caption{\textbf{Left panels:} From top to bottom, the UV and X-ray flux for the Fiducial, Low Flux, Power-Law, and Shallow Power-Law runs are shown in blue and orange, respectively, while the UV flux from the Gray run without injected X-rays is shown in black for comparison. \textbf{Right panels:} The time- and horizontally-averaged vertical profiles for density (top panel), gas temperature (middle panel), and turbulent velocity (bottom panel) for the four runs with injected X-rays (in color), and for the Gray run without injected X-rays (dashed black line).}
    \label{fig:ch5_disk}
\end{figure*}

\begin{figure}
    \includegraphics[width=0.49\textwidth]{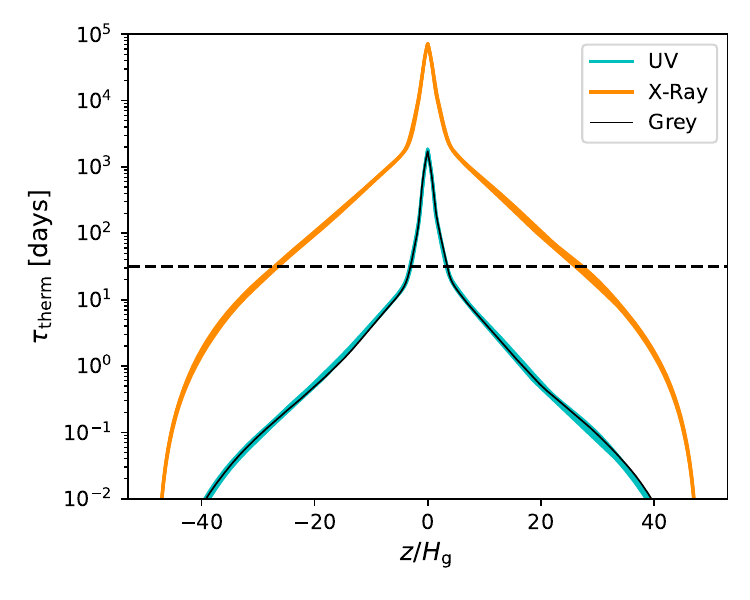}
    \caption{The horizontally- and time-averaged vertical profile for the UV and X-ray thermal timescales for all runs are shown in blue and orange, respectively. The solid black line shows the vertical profile for the UV thermal timescale from the Gray run without injected X-rays. The dashed black line shows the average thermal timescale for the disk for the Fiducial model.}
    \label{fig:ch5_ttherm}
\end{figure}

Before presenting the light curves from our simulations, in this section we briefly describe the structure of the disk in our four runs. We compare the X-ray and UV flux in all four runs to the UV flux in our Gray run (black line) in the left panel of Figure \ref{fig:ch5_disk}. For all runs, the X-ray flux makes it less than a scale height into the disk. Instead, at the surface of the disk the UV flux is enhanced by roughly the amount of X-ray flux entering the simulation. Beneath the surface, however, there is less UV flux than in the Gray run. This decreased UV flux is most significant for the Fiducial run and least significant for the Shallow Power-Law run, suggesting that the closer the PSD of the X-ray irradiation is to a DRW the more the X-rays suppress UV emission below the surface of the disk. The change in UV flux may be due to enhanced turbulent dissipation or convection in these runs.

In the right panel of Figure \ref{fig:ch5_disk} we show the time- and horizontally-averaged vertical profiles for the density, gas temperature, and turbulent velocity. The properties of the disk are qualitatively consistent between all four runs. The gas density stays consistent for runs with and without injected X-rays (our Gray run shown with the dashed black line). For the turbulent velocity, the largest difference between our runs with and without injected X-rays is a larger spike in the velocity between $z=\pm 2.5-8$~$H_{\rm g}$. This region is where the magnetic pressure increases and becomes the dominant source of pressure in our simulated disks. However, the temperature of the disk is slightly lower in this region for the runs with injected X-rays, which increases the ratio of the magnetic pressure to the gas and radiation pressure significantly. As a result, there is greater turbulence in this region in our irradiated disks.

The biggest difference in these vertical profiles between our Gray run and our runs with injected X-rays is a factor of 2--5 increase in the gas temperature beyond $\pm30 H_{\rm g}$. While the X-rays do not penetrate down to $\pm30 H_{\rm g}$, they do provide a force that compresses the gas near the surface of the disk. Because this force varies with the X-ray flux, it provides a source of fluctuating mechanical power that heats the disk over $10$~$H_{\rm g}$ below where the X-rays penetrate. The UV opacity in this region of the disk is scattering dominated and the disk is not thermalized, so despite the significant increase in temperature, the UV flux from this region does not increase. Nevertheless, this temperature increase could modify the spectrum emitted from this region by producing more high energy photons than will be produced from a disk without irradiation.


We show the time- and horizontally-averaged vertical profile for the thermal timescale, $\tau_{\rm therm}(z)=\tau_{\rm UV,XR} L/c$, where $L$ is the distance to the edge of the simulation box and $\tau_{\rm UV,XR}$ is the UV or X-ray optical depth, in Figure \ref{fig:ch5_ttherm}. The dashed black line shows the time-averaged thermal timescale $\langle \tau_{\rm therm} \rangle=1/(\langle \alpha\rangle \Omega)$, where $\alpha$ is the \cite{Shakura1973} viscosity parameter and $\Omega$ is the orbital frequency, for the Fiducial run. The thermal timescales are consistent between all four runs with injected X-rays as well as the Gray run from \citetalias{Secunda:2024} without injected X-rays (black solid line). The thermal timescale of our simulated disk is important because it sets the timescale of fluctuations in the disk. There is also observational evidence that the thermal timescale sets the damping timescale for DRW-modeled UV-optical light curves emitted by AGN disks \citep{Burke2021}.

\subsection{Light Curves}
\label{sec:ch5_lcs}

\begin{figure*}
    \includegraphics[width=0.5\textwidth]{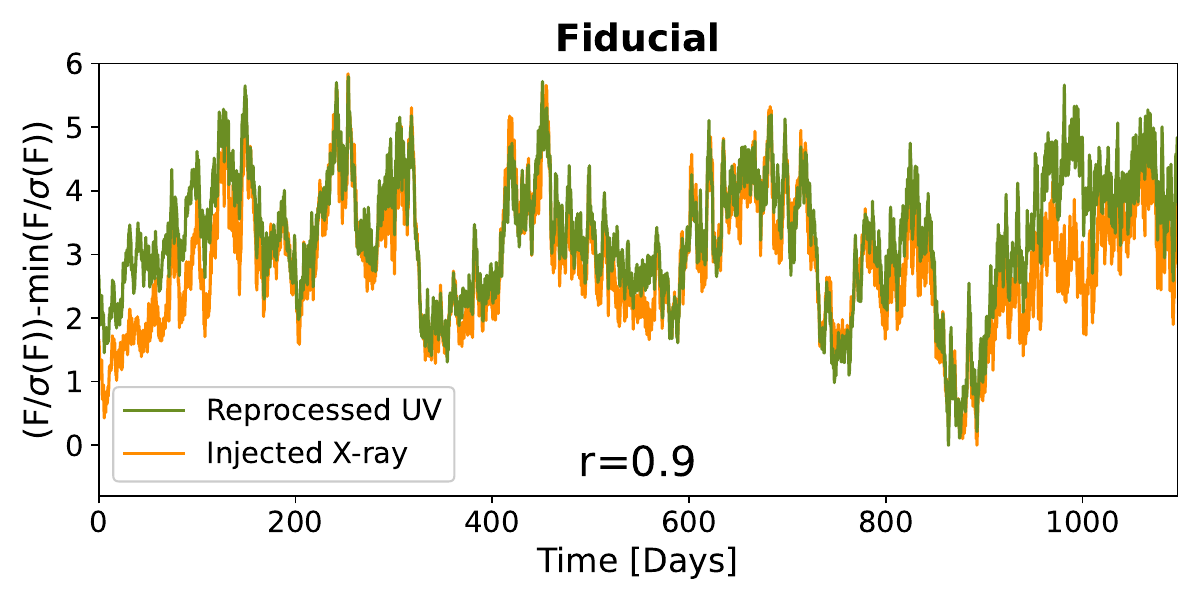} 
    \includegraphics[width=0.5\textwidth]{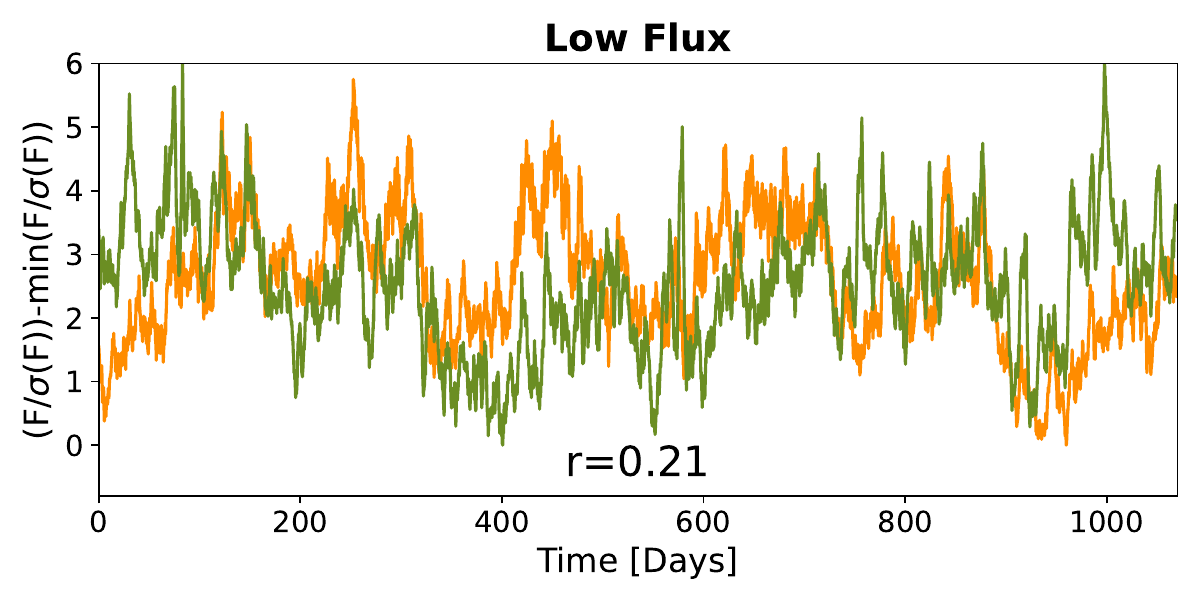} \\
    \includegraphics[width=0.48\textwidth]{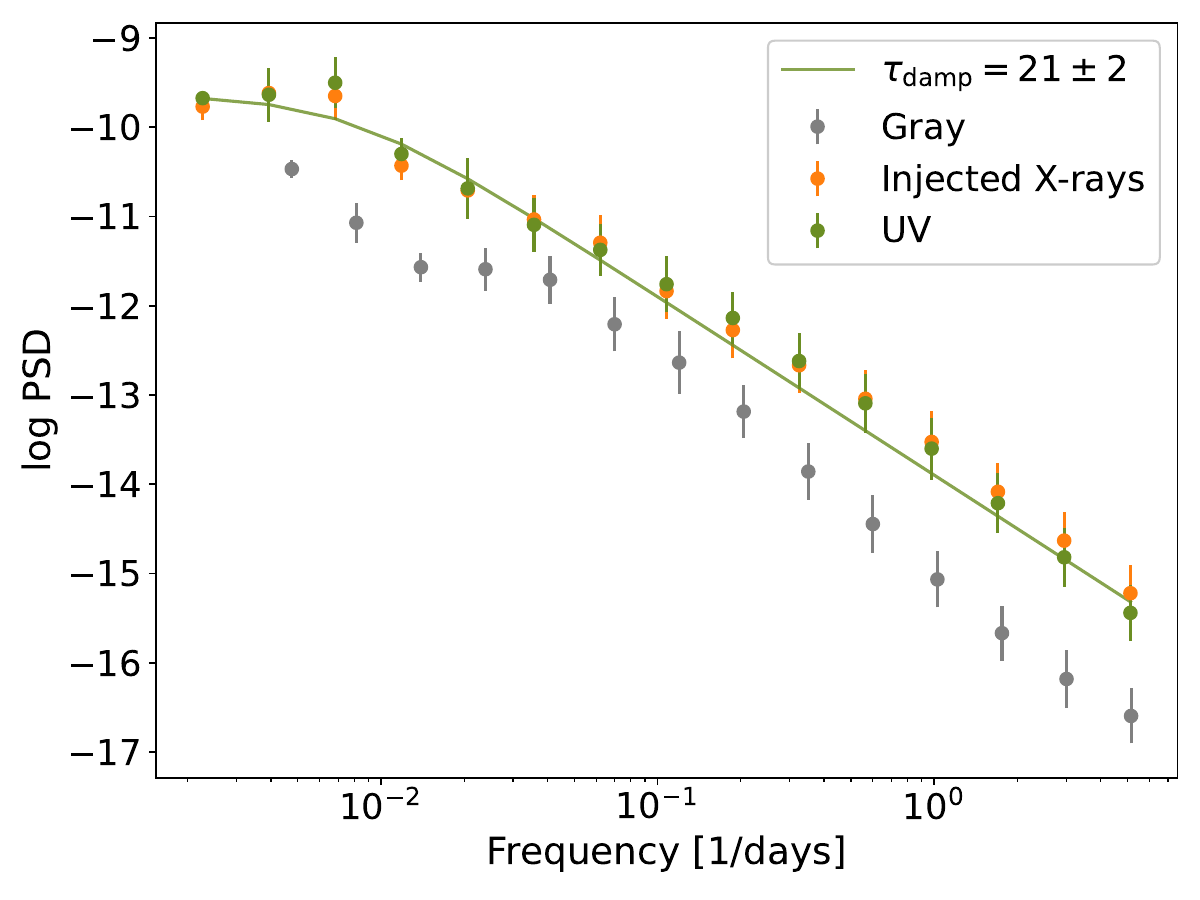} 
    \includegraphics[width=0.48\textwidth]{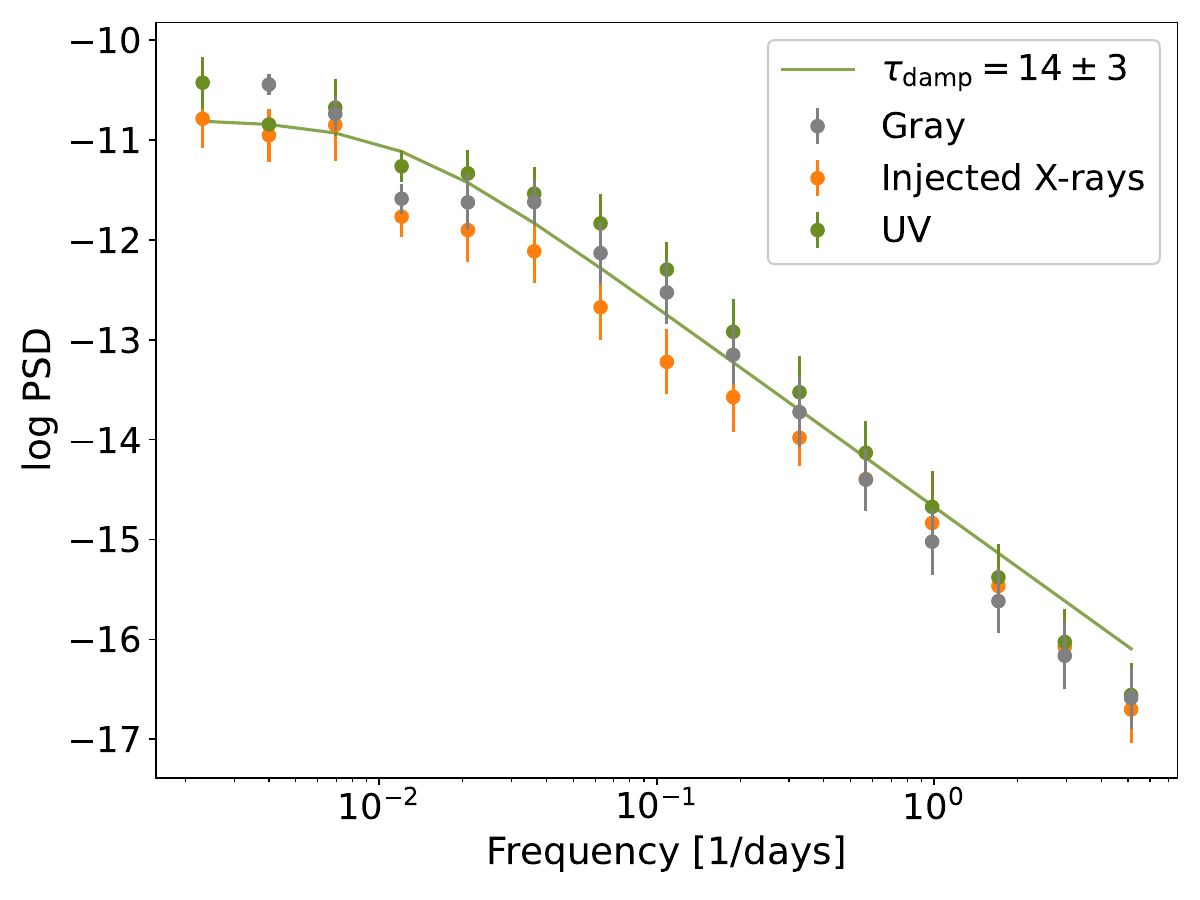}  \\
    \includegraphics[width=0.5\textwidth]{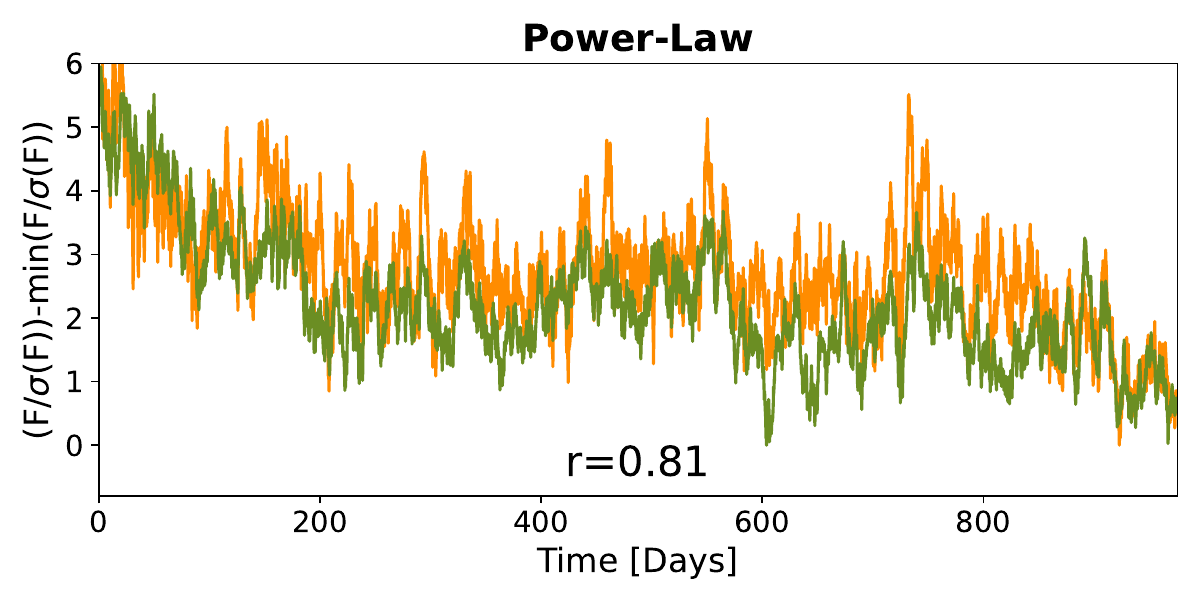}
    \includegraphics[width=0.5\textwidth]{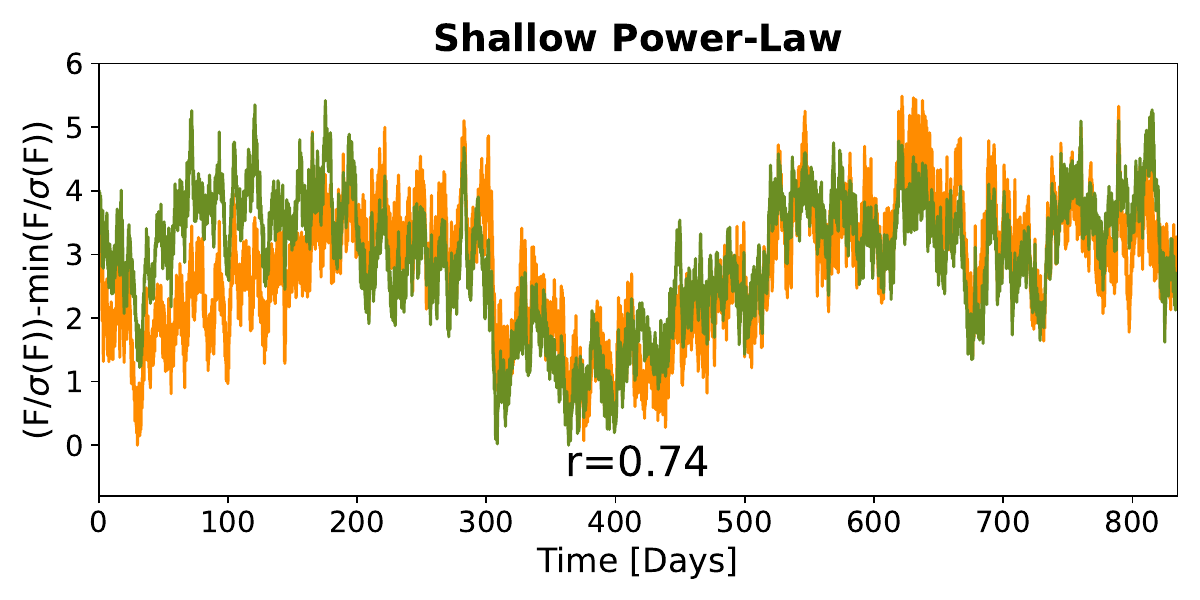} \\
    \includegraphics[width=0.48\textwidth]{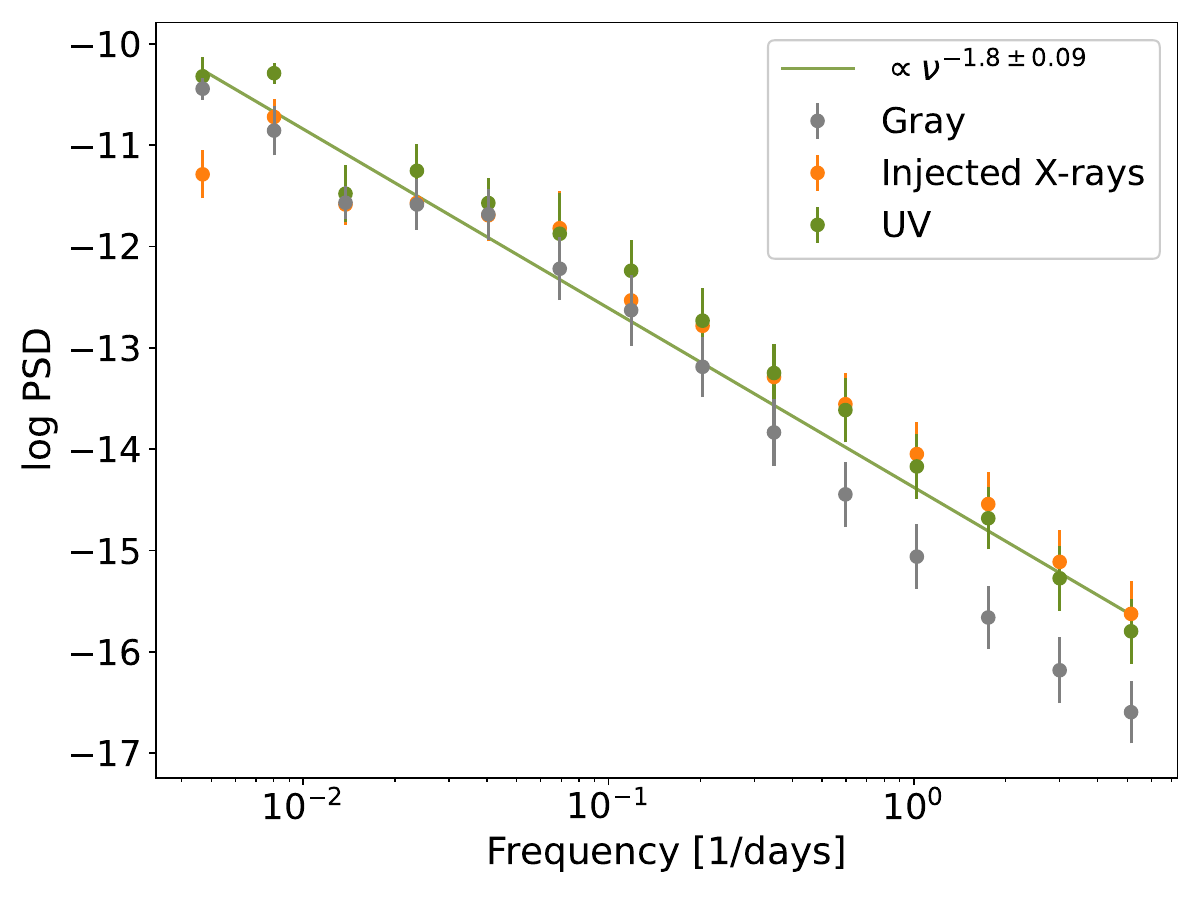} 
    \includegraphics[width=0.48\textwidth]{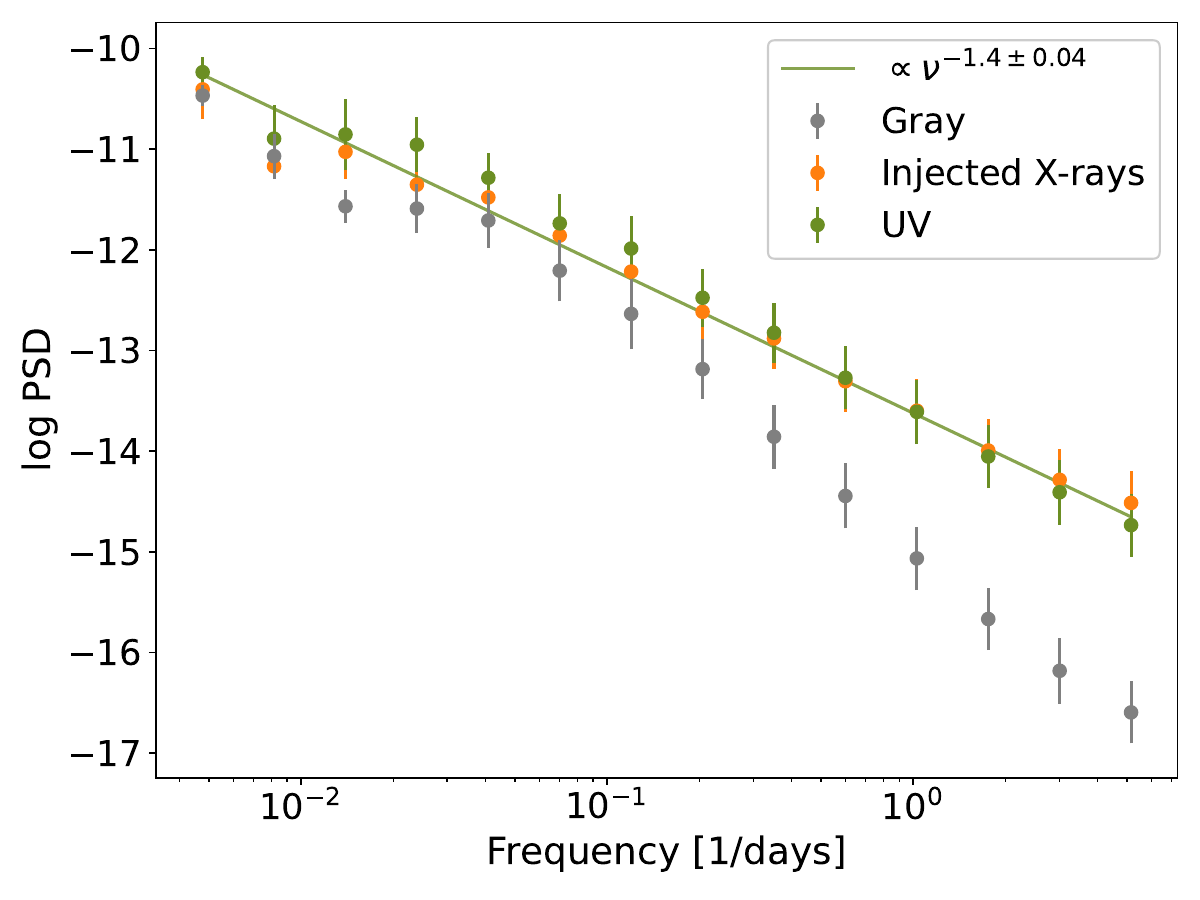}
    \caption{\textbf{Upper panels:} Injected X-ray (in orange) and reprocessed UV (in green) light curves from our four different runs, labeled with the Pearson r-correlation coefficient between the two light curves. For easier comparison, we scale the flux, $F$, by the standard deviation of the flux, $\sigma(F)$, and subtract off the minimum flux for each light curve. We inject the X-rays at time equals zero. Note, because X-rays are injected directly into the simulation there will be no lag between the X-ray and UV light curves from the light travel time. \textbf{Lower panels:} PSDs for the injected X-ray, reprocessed UV, and intrinsic UV (from the Gray run with no injected X-rays) light curves from our four runs. We show best fit DRW-models for the reprocessed UV PSDs for the Fiducial and Low Flux runs and best fit power-law models for the reprocessed UV PSDs from the two power-law runs.}
    \label{fig:ch5_lcs}
\end{figure*}

We present the simulated light curves from our four runs in the upper panels of Figure \ref{fig:ch5_lcs}. The fluxes have all been normalized for easier comparison. The lower panels of Figure \ref{fig:ch5_lcs} show the PSDs for these light curves as well as the PSD of the intrinsic UV light curve from the Gray run in \citetalias{Secunda:2024} with no injected X-rays for comparison.

In our Fiducial run, the injected X-ray and the emitted UV light curves are highly correlated, which is clear both by eye in the top left panels of Figure \ref{fig:ch5_lcs} and from the Pearson r-coefficient between them of $r=0.90$.  In addition, the PSD of the reprocessed UV light curve is mostly consistent with the X-ray PSD, except at the highest frequencies where the reprocessed UV has slightly less power. The intrinsic UV PSD has less power at all frequencies. It is clear the variability in the UV light curve is nearly entirely from the reprocessing of the X-ray variability, despite the UV light curve from our Gray run having its own intrinsic variability. We use the least squares method to fit a DRW model to the reprocessed UV PSD from our Fiducial run and find $\tau_{\rm damp}=21 \pm 2$~days, which unsurprisingly is similar to the damping timescale used to model the injected X-ray light curve (30~days).

In our Low Flux run (top right panels of Figure \ref{fig:ch5_lcs}), we inject the same X-ray light curve as in our Fiducial run, but with 1/4 the X-ray flux. Decreasing the X-ray flux leads to a much lower correlation between the X-ray and reprocessed UV light curves ($r=0.21$). Unlike in the Fiducial run, the PSDs of the reprocessed and intrinsic UV light curves from the runs with and without injected X-rays, respectively, are more similar than the injected X-ray and reprocessed UV PSDs. In particular, both UV PSDs have more power at $f<0.3$~days$^{-1}$ than the injected X-ray PSD and a best-fit damping timescale of $\tau_{\rm damp}=14 \pm 3$~days. Long-baseline optical AGN light curves suggest that the damping timescale of DRW light curves roughly corresponds to the thermal timescale of AGN disks \citep{Kelly2009,Burke2021,Stone:2022}. The damping timescale for this Low Flux run, where most of the variability appears to be driven by fluctuations in the AGN disk instead of by X-ray variability, is roughly half the average thermal timescale in Figure \ref{fig:ch5_ttherm} of about 30~days.

Our two runs in which we model the X-ray PSD as a power-law, the Power-Law ($\propto \nu^{-2}$) and Shallow Power-Law ($\propto \nu^{-1.5}$) runs, are shown in the bottom panels of Figure \ref{fig:ch5_lcs}. The X-ray and reprocessed UV light curves are well-correlated for the Power-Law and Shallow Power-Law runs, with $r=0.81$ and $r=0.74$, respectively. The X-ray and UV PSDs are very similar and differ from the intrinsic UV PSD at high frequencies, where the intrinsic UV PSD has significantly less power. The difference between the intrinsic and reprocessed UV PSDs is greater for the Shallow Power-Law run due to the shallower slope of the injected X-ray PSD. We fit a power-law to the reprocessed UV PSD for both of these runs and find that they are well-fit by power-laws slightly shallower than the power-law used to generate the injected X-ray light curve. Overall, changing the PSD of the injected X-ray light curve has some effect on the correlations between the different light curves, but lowering the amount of flux in the injected X-ray light curve seems to have the largest effect.

\subsection{Response Functions}
\label{sec:ch5_response}

\begin{figure*}
    \centering
    \includegraphics[width=0.32\textwidth]{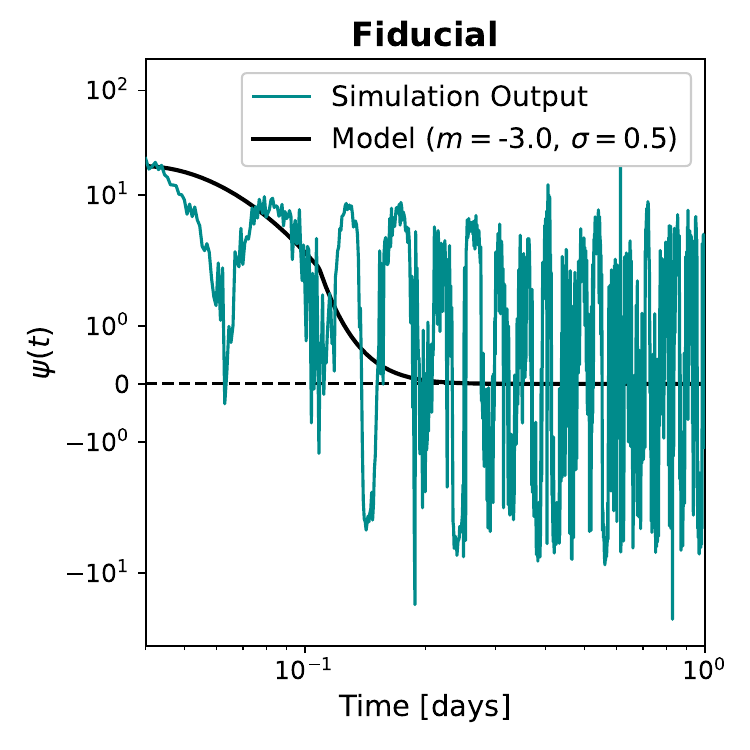} 
    \includegraphics[width=0.32\textwidth]{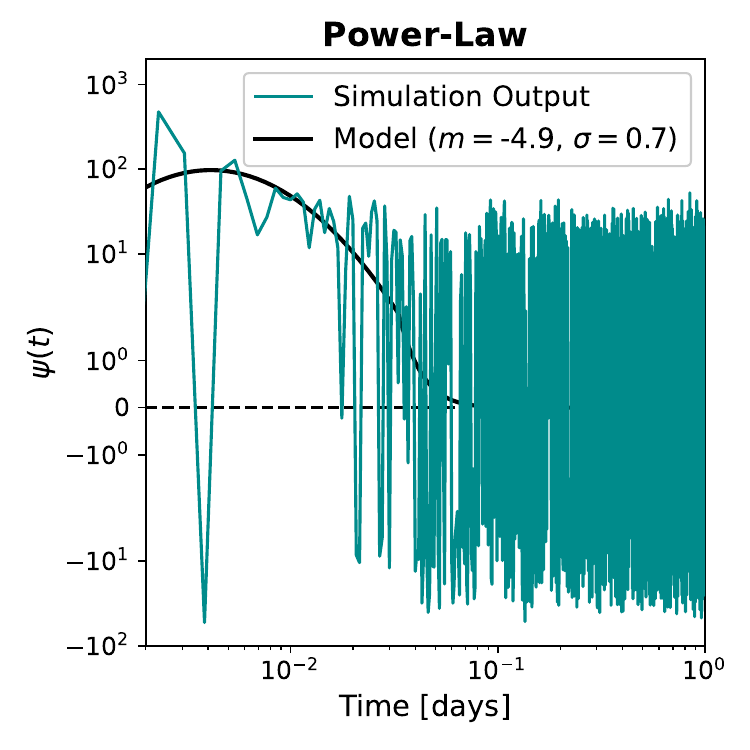} 
    \includegraphics[width=0.32\textwidth]{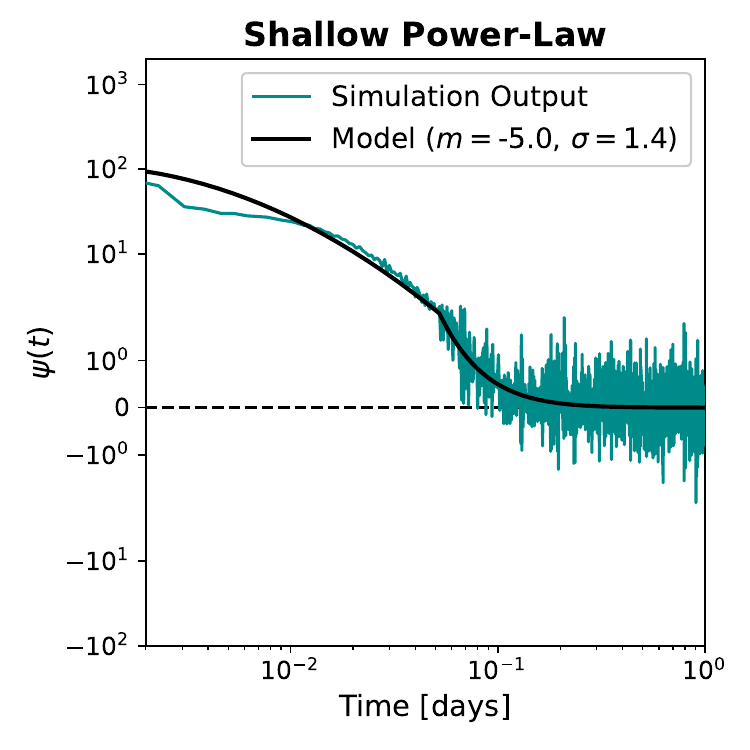} 
    \\
    \includegraphics[width=0.32\textwidth]{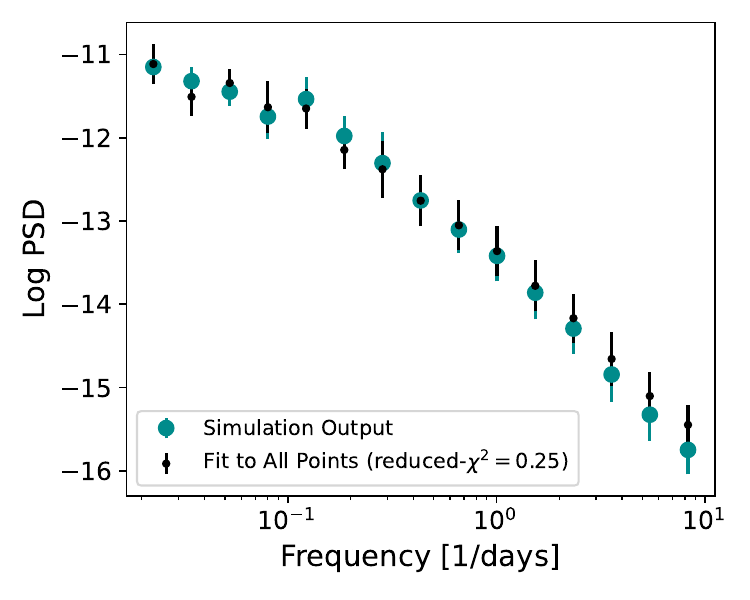} 
    \includegraphics[width=0.32\textwidth]{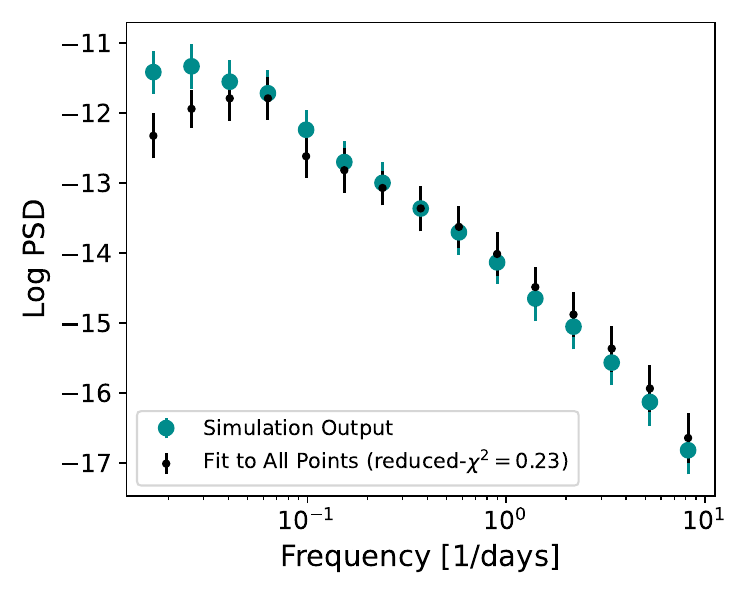} 
    \includegraphics[width=0.32\textwidth]{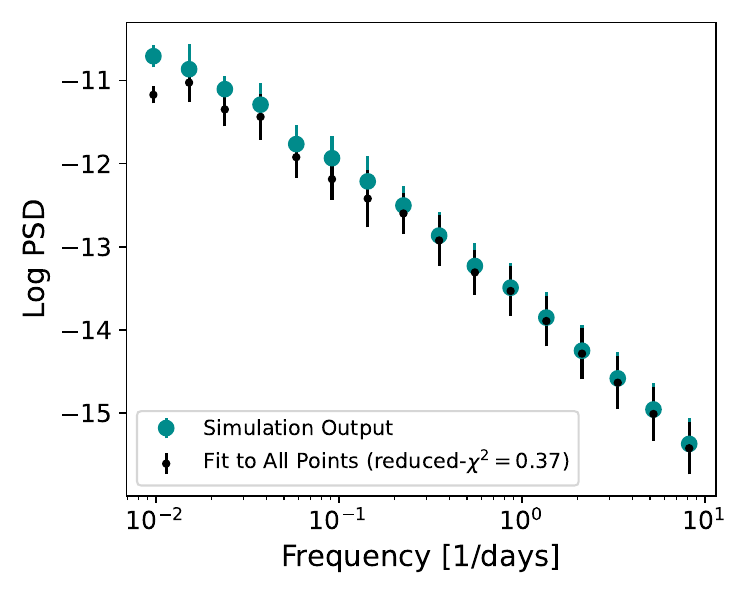} \\
    \includegraphics[width=0.32\textwidth]{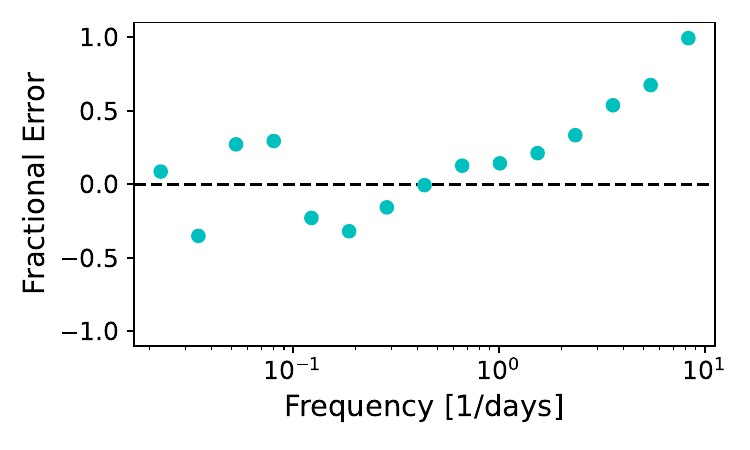} 
    \includegraphics[width=0.32\textwidth]{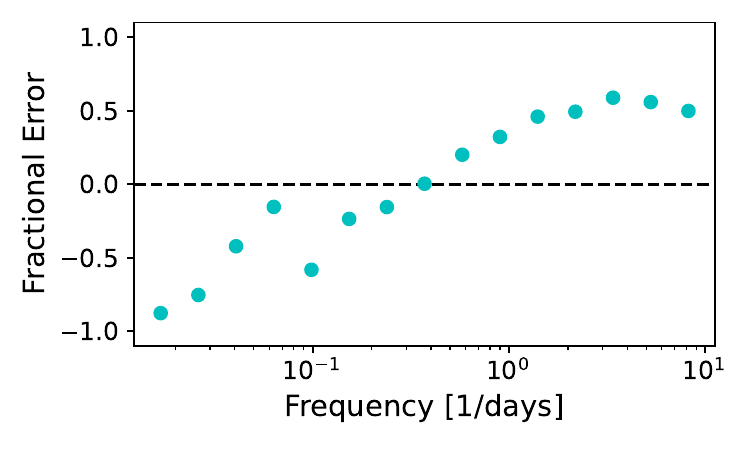} 
    \includegraphics[width=0.32\textwidth]{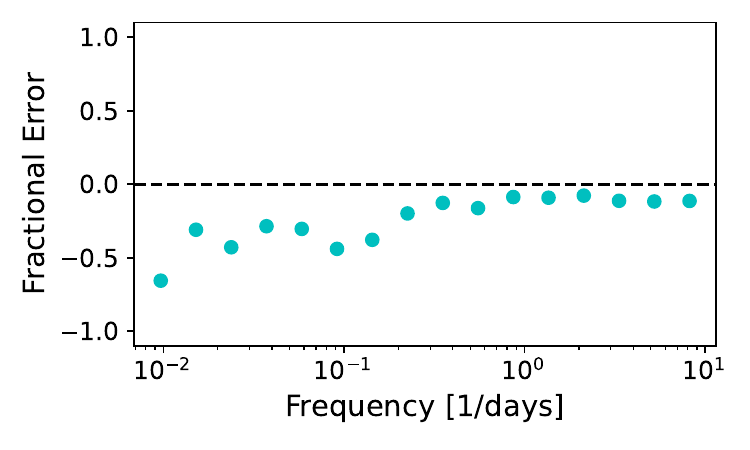} 
    \caption{\textbf{Top panels:} The response function in real space as a function of time, $\psi(t)$, calculated using our simulation data (in green). The black line shows our best-fit log-normal models for the simulation output. $m$ and $\sigma$ for Equation \ref{eq:lognorm} are given in the legend for each run. \textbf{Middle panels:} The logarithm of the binned PSD for the simulation data is shown in green and the logarithm of the binned PSD of the X-ray light curve reprocessed with the response function shown in the upper panels in black. Reduced-$\chi^2$ for each model is given in the legend. \textbf{Bottom panels:} The fractional error, (modeled PSD - simulation output PSD)/(simulation output PSD), as a function of frequency. Note we do not fit a response function to our Low Flux run because of the low correlation between the X-ray and UV light curves.}
    \label{fig:fid_poly}
\end{figure*}

We can use the well-correlated X-ray and UV light curves from the Fiducial, Power-Law, and Shallow Power-Law runs presented in the previous section to test the assumptions of the linear reprocessing model commonly used in reverberation mapping. The reprocessing of X-ray emission from the corona by the AGN disk into UV-optical light is modeled using a response function, $\psi(t)$, which is typically assumed to be a Gaussian or log-normal function. Because our simulations produce evenly sampled driving X-ray and reprocessed UV light curves we can fast Fourier transform these light curves $r(t) \rightarrow R(\nu)$ and $d(t) \rightarrow D(\nu)$ and use the Fourier transform of Equation \ref{eq:ch5_reproccess}, 
\begin{equation}
\label{eq:ch5_freprocess}
        \rm R(\nu) = \Psi(\nu)D(\nu),
\end{equation}  
to solve for the Fourier transform of the response function, $\Psi(\nu)$. We then inverse Fourier transform $\Psi(\nu) \rightarrow \psi(t)$ to derive the response function as a function of time in real space. We show these response functions in the top panels of Figure \ref{fig:fid_poly}. 

We experimented with several functional forms for a model to fit to this raw data. We find that the best-fitting function is the often used log-normal response function,
\begin{equation}
    \label{eq:lognorm}
    \Psi(t)= \frac{1}{t\sigma \sqrt{2\pi}}\exp{\left[-\frac{(\ln{t} - m)^2}{2\sigma^2}\right]}.
\end{equation}
We show the best-fitting log-normal response functions in black in the top panels of Figure \ref{fig:fid_poly}. 

$m \ll 0$ for both power-law runs, meaning reprocessing is nearly instantaneous. Even for the Fiducial run, where $m=-3.0$, reprocessing would only be delayed on hour-long timescales, which would not have a significant impact on measured lag durations. In addition, for all three runs the response functions are narrow, with $\sigma<e$, suggesting there is not a lot of smoothing or smearing out of the signal over time as the X-ray irradiation interacts with the gas disk. Overall, these response functions suggest that reprocessing occurs on timescales of two hours or less, which is consistent with the assumptions typically made in most reprocessing models \citep[e.g.,][]{Cackett:2007}.

To check the fit of our modeled response function we convolve the X-ray light curve with these best-fit models and compare this forward-modeled PSD to the actual UV PSD in the middle panels of Figure \ref{fig:fid_poly}. The fractional error, (modeled PSD - simulation output PSD)/(simulation output PSD) for these PSDs is shown in the bottom panel of Figure \ref{fig:fid_poly}. We give the reduced-$\chi^2$ value for these fits in the legend. For all runs, reduced-$\chi^2<1$ due to the large variance in the PSD of the reprocessed UV light curves we fit our models to. 

Despite these low reduced-$\chi^2$ values, our best-fit models for the Fiducial and Power-Law runs over-estimate the amount of variability at high frequencies by up to a factor of 2, failing to capture how disk reprocessing in our simulations smooths away some of the high frequency variability. On the other hand, our best-fit models for the Power-Law and Shallow Power-Law runs underestimate the amount of variability at low frequencies by a similar margin. This underestimate is likely due to the fact that there is more power at low frequencies in the reprocessed UV PSD than the X-ray PSD (see bottom panels of Figure \ref{fig:ch5_lcs}), which suggests that reprocessed X-ray irradiation is not the only driver of low frequency variability in the UV light curve. This additional low frequency variability may be the reason why the X-ray and UV light curves from these power-law runs are less correlated than in the Fiducial run. These results indicate that a higher-order model where there is a different ratio of reprocessing at different frequencies may be a better fit to our simulation data than a linear reprocessing model.

\subsection{Lag Detection}
\label{sec:ch5_javelin}

\begin{figure}
    \centering
    \includegraphics[width=0.5\textwidth]{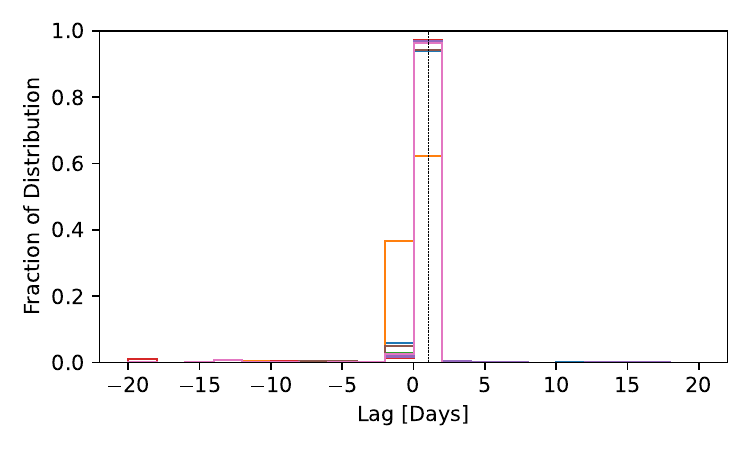} \\
    \includegraphics[width=0.5\textwidth]{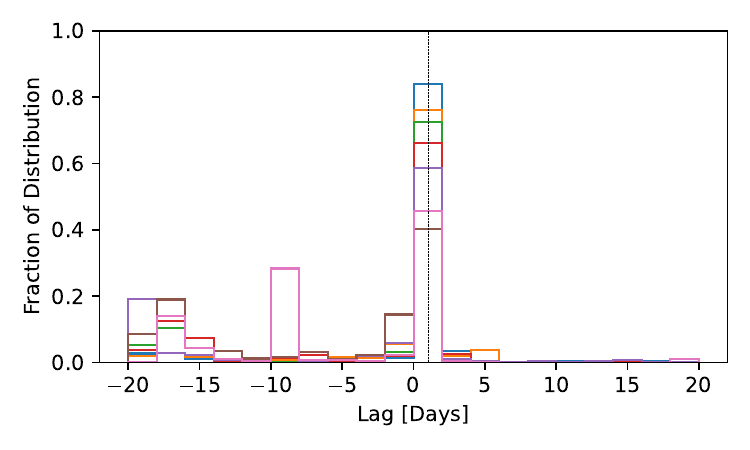}
    \caption{{\sc javelin} distribution for the lags detected in seven mock light curves made using light curves from the Fiducial run (top panel) and Shallow Power-Law run (bottom panel). The vertical black line shows the one day lag we added by shifting the X-ray light curves forward one day.}
    \label{fig:jav}
\end{figure}

Our simulated light curves self-consistently reprocess our injected X-ray light curves into well-correlated UV light curves for three out of our four runs. As a result, if we artificially lag our UV light curves by a day we should be able to recover this one day lag with commonly used lag recover methods. Many lag recovery methods such as {\sc javelin} \citep{Zu2011,Zu2016}, {\sc cream} \citep{Starkey:2016}, Gaussian process regression \citep[e.g.,][]{Lewin:2023}, and machine learning methods \citep[e.g.,][]{Fagin:024,Li:2024}, assume light curves can be modeled as DRWs and make assumptions about the response function for the reprocessing. These methods are then tested on mock light curves with similar to identical assumptions. We encourage the community to use our publicly available mock light curves (see footnote in Section \ref{ch5:sec:discuss}) to test these lag recovery methods, because they include rarely modeled elements such as fluctuation-driven intrinsic variability and reprocessing without an assumed response function.

As an example, in this section we test the lag recovery method {\sc javelin} on our light curves. {\sc javelin} simultaneously models the driving light curves as a DRW while also fitting the lag between the driving and reprocessed light curves using a top hat function centered at the mean lag. The light curves simulated here as well as their response functions are a significant departure from the way they will be modeled by {\sc javelin}, and so they should be a good test for the efficacy of this recovery method.

To make mock light curves to feed into {\sc javelin} we take our X-ray light curve and shift it forwards by one day. We then divide our light curves up into seven $\sim 300$~day segments and sub-sample the X-ray and UV light curve with the observing cadence for two different wavebands from the first half of the {\sc swift} observing campaign of Fairall 9 \citep{F92020}.

We show the {\sc javelin} distributions for the lags detected in these seven mock light curves for our Fiducial and Shallow Power-Law runs in the top and bottom panels of Figure \ref{fig:jav}, respectively. The {\sc javelin} distributions for all mock light curves peak around the input lag of one day. The distributions peak more strongly around one day for the Fiducial run light curves, which may due to the higher correlation between the two light curves or the fact that the injected X-ray light curve in this case is modeled as a DRW. Nevertheless, {\sc javelin} is still accurate in detecting the lag between the X-ray and UV light curves from the Shallow Power-Law run, despite the PSD used to generate the X-ray light curve in this simulation being significantly different from a DRW. We encourage others to use our publicly available light curves (see footnote in Section \ref{ch5:sec:discuss}) to test software that has been developed or trained assuming other light curve and reprocessing models.

\section{Discussion}
\label{ch5:sec:discuss}

In \citetalias{Secunda:2024}, we used the multi-frequency radiation MHD code described in \cite{Jiang2022} to simulate the reprocessing of high-frequency X-rays into lower-frequency UV light by the gas disk. This reprocessing is a key but understudied component of disk continuum reverberation mapping, one of the primary tools used to learn about AGN disks. We found that the X-ray and UV light curves were not strongly correlated due to very low X-ray absorption opacities, making it apparent that it is necessary to use sophisticated opacity models that include line opacities for the X-ray opacities in addition to the UV opacities. In order to investigate further, in this paper we use the tabulated opacities from {\sc tops} that include these line opacities for our X-ray radiation. We find that depending on the amount of flux and the PSD of the variability of the injected X-ray light curve, the X-ray irradiation and reprocessed UV light curve can be highly correlated with reprocessing occurring on hour timescales. In the near future, it may be possible to observe line emission using X-ray spectra, which would allow us to further test X-ray opacity models.

Our Fiducial run here is identical to run A in \citetalias{Secunda:2024}, except for the updated opacity model for the X-ray radiation. Instead of the weak correlation of $r=0.34$ found in \citetalias{Secunda:2024}, we find a strong correlation of $r=0.90$ between the X-ray and UV light curves. This stronger correlation is due to the greater absorption opacity for the X-ray photons which can allow for an energy transfer from the X-ray irradiation to the gas, instead of only the momentum transfer we observed in \citetalias{Secunda:2024}. 

We also find that while the X-ray irradiation makes it less than a scale height into the disk, the X-rays still heat the gas down to around $\pm 30$~$H_{\rm g}$ due to the impact force of the irradiation which mechanically heats the gas. As a result, temperatures in the optically thin region can be a factor of 2--5 larger than in our run without X-ray irradiation. This region of the disk is not thermalized, and so the UV emission does not increase by a large factor. However, these higher temperatures could 
modify the disk spectrum and produce more high energy photons at large radii than will be produced from a disk without irradiation. This result can help explain recent reverberation mapping campaigns that find that the radial extent of the disk is 2--3 times larger than anticipated \citep[e.g.,][]{Jiang:2017,Guo:2022}. Future work should do a more careful calculation of the full spectra emitted from this region of the disk, which may not be well represented by a single temperature Blackbody.

In addition to our Fiducial run, we perform three different runs where we vary the amount of injected X-ray flux and the PSDs of the injected X-ray light curve. Changing the PSD of the injected X-ray light curve from a DRW, as in our Fiducial run, to a power-law with index $-2$ and $-1.5$ for our Power-Law and Shallow Power-Law runs, respectively, leads to a slightly lower correlation between our X-ray and reprocessed UV light curves. On the other hand, decreasing the flux by a factor of 4 in our Low Flux run leads to a significantly lower correlation between the X-ray and reprocessed UV light curves. Since varying the mean X-ray flux has a large impact on the correlation between the X-ray and UV light curves, the observed weak to moderate correlations for several AGN could result from changing X-ray flux levels \citep[e.g.,][]{Edelson:2019,Cackett:2023,Kara:2023,Schimoia:2015,Buisson:2018}. For all but the Low Flux run, how we model the PSD of the X-ray light curve determines the PSD of the reprocessed UV light curves because of the high correlation between the two light curves. For the Low-Flux run we fit a DRW model to the UV PSD and find a damping timescale of $\tau_{\rm damp}=14\pm3$~days, about half the average thermal timescale of our simulated disks, which is roughly consistent with observational evidence that the thermal timescale sets the damping timescale of UV-optical AGN light curves \citep{Kelly2009,Burke2021,Stone:2022}.


We fit log-Normal functions to the response functions derived from our simulations, as they are often used to fit observational data. Our best-fit response functions are narrow and centered at timescales of one hour or shorter, which agrees with the typical assumptions made that disk reprocessing is instantaneous, i.e., does not contribute to the lag duration \citep{Cackett:2007}. However, these response functions are not good fits at all frequencies, suggesting that a higher-order reprocessing model where there is a different ratio of reprocessing at different frequencies may be a better fit to our simulation data than a linear reprocessing model.

The popular lag detection method {\sc javelin} is able to accurately recover one-day lags that we add to mock light curves we make by sub-sampling our simulated light curves with the cadence from the Fairall 9 {\sc swift} campaign \citep{F92020}. The accuracy of {\sc javelin} is reassuring, given we do not model our light curves as a DRW reprocessed with a simple top hat function as {\sc javelin} expects. In fact, the PSDs of the UV and X-ray light curves in the Shallow Power-Law case are a significant departure from a DRW. The light curves from our simulations are publicly available\footnote{Light curves available at: \url{https://github.com/AmySec/RMHD_lightcurves}} and could be powerful tools for testing different lag detection software that has been developed or trained based on more traditionally modeled light curves that do not self-consistently incorporate the physics of the turbulent gas disk.

One shortcoming of our simulations is that they are local approximations, or shearing boxes. If the whole disk can be treated as the sum of many shearing boxes, then these shearing boxes will have random phases relative to each other, but the variability timescales will remain the same. As a result, using $n$ shearing boxes would shift the PSD of the variability to lower power, as demonstrated analytically by \citet{Dexter:2011}, reducing the PSD by a factor of $1/\sqrt{n}$. If we assume a constant radius and arrange shearing boxes along the azimuthal direction, the X-ray variability would remain coherent while the UV PSD would decrease, potentially resulting in a higher correlation between the injected X-ray and emitted UV light curves. For example, for the Low Flux run where the X-ray and UV light curves are poorly correlated, the UV PSD is on average a factor of $\sim 3$ greater than the injected X-ray PSD. Therefore, future work could use $n=9$ shearing boxes in the azimuthal direction to see if a larger simulation region would increase the correlation of X-ray and UV light curves. However, \citet{Ren:2024} showed that even when accounting for inhomogeneous DRW variability along the azimuthal direction, disk fluctuations can decrease the correlation between X-ray and UV light curves.


Ultimately, expanding these multi-frequency shearing box simulations to global simulations of the UV-optical emitting region of the AGN disk is necessary to fully understand the length-scales for the coherence of disk fluctuations, and how incoherence between different azimuthal or radial regions impacts the UV PSD and the correlation between X-ray and UV light curves. Additionally, with global simulations we can study longer wavelength/lower mode fluctuations that can not be captured by local shearing box simulations, and how all of these fluctuations might be propagated inward or outward through the disk, impacting the variability of light curves emitted in other regions of the disk. Evidence for the propagation of disk fluctuations have been observed by \citet{Neustadt:2022}, \citet{Yao:2022}, and others and could provide information on the vertical structure of AGN disks \citep{Secunda2023}. With global simulations we can also examine the impact of the angle at which the disk is irradiated, allowing us to probe the open question of corona geometry. Finally, we see in our shearing box simulations that X-ray irradiation increases the temperature of the disk in the optically thin region, and global simulations will help us better understand the impact of X-ray irradiation on the AGN disk and the spectrum emitted by the disk. 

Updating the way we model reverberation mapping is crucial to prepare for the Vera Rubin Observatory, which will carry out a comprehensive survey of the southern sky and contain over 100 million quasars \citep{Ivezi2019}. Reverberation mapping campaigns often compare their empirical results to the \cite{Shakura1973} optically thick but geometrically thin standard disk model. However, there is growing evidence that the standard thin disk model does not fully describe an AGN disk \citep[e.g.,][]{Antonucci:1988,Antonucci:1989,Antonucci:2023}. For example, both recent simulations and observations of AGN disks suggest that pressure support from magnetic fields and radiation can lead to puffed up or slim instead of thin disks \citep[e.g.,][]{Gaburov:2012,Jiang2016,Jiang:2019,JiangBlaes2020,Hopkins:2023,Yao:2022,Secunda2023}. In addition, both micro-lensing and reverberation mapping campaigns suggest that the radial extent of the disk is larger than predicted by the standard disk model \citep[e.g.,][]{Morgan:2010,Jiang:2017,Guo:2022}. Understanding the physical processes behind reprocessing and AGN light curve variability along with the broad data set for Rubin will allow us to develop new models for AGN disks that go beyond the standard thin disk model.

\begin{acknowledgements}
    The authors would like to thank Dr. Tim Kallman for useful discussion on X-ray opacities and the referee for their thoughtful comments. JEG is supported in part by NSF grants AST1007052, AST1007094, and 22624. The Center for Computational Astrophysics at the Flatiron Institute is supported by the Simons Foundation. 
\end{acknowledgements}

\bibliography{agn_lag.bib}
\end{CJK*}
\end{document}